%% file: arxiv_submission_190808.tex
\documentclass[%
 reprint,
 superscriptaddress,
 nofootinbib,
 amsmath,amssymb,
 aps,
 pra,
]{revtex4-1}
 
\usepackage{amsmath}
\usepackage{amssymb}
\usepackage[caption=false]{subfig}
\usepackage{graphicx}
\usepackage[utf8]{inputenc}
\usepackage[T1]{fontenc}
\usepackage[english]{babel}
\usepackage{xcolor}

\newcommand{\p}{\partial}
\newcommand{\avg}[1]{\left\langle #1 \right\rangle}

\newcommand{\tw}{\tilde{w}}
\newcommand{\ha}{\hat{a}}
\let\Re\relax
\DeclareMathOperator{\Re}{Re}
\let\Im\relax
\DeclareMathOperator{\Im}{Im}
\DeclareMathOperator{\acosh}{acosh}
\newcommand{\Stunn}{S_\mathrm{tunn}}

\graphicspath{{../../figures/}}

\begin{document}
\title{Effects of tunneling and multiphoton transitions on squeezed--state
       generation in bistable driven systems}
\date{\today}
\author{Natalya S. Maslova} 
\affiliation{
    Department of Quantum electronics and Quantum Center, 
                        Faculty of Physics, Moscow State University, 119991 Moscow, Russia
    }
\author{Evgeny V. Anikin}
\affiliation{Skolkovo Institute of Science and Technology, 121205 Moscow, Russia}
\affiliation{Moscow Institute of Physics and Technology, 141701 Dolgoprudny, Russia}
\author{Nikolay A. Gippius}
\affiliation{Skolkovo Institute of Science and Technology, 121205 Moscow, Russia}
\author{Igor M. Sokolov}
\affiliation{Institut für Physik and IRIS Adlershof, Humboldt Universität 
             zu Berlin, Newtonstraße 15, 12489 Berlin, Germany}

\begin{abstract}
Bistability of a nonlinear resonantly--driven oscillator in the presence of external noise 
is analyzed using the classical Fokker--Planck equation in the quasienergy space 
with account for tunneling effects
and by quantum master equation in quasienergy states representation. Two timescales 
responsible for different stages of this bistable system relaxation have been obtained. 
We found that the slow relaxation rate caused by fluctuation--induced 
transitions between different stable states can be enhanced by several orders of magnitude 
due to the tunneling effects. It was also revealed that tunneling between nearly 
degenerate quasienergy states and resonant multiphoton transitions between the 
genuine eigenstates of the nonlinear oscillator are just the similar effects.
It was demonstrated that the quasienergy states in the
bistability region corresponding to
higher amplitude are squeezed. The degree of squeezing is determined 
by the ratio between nonlinearity and 
detuning, so that the uncertainty of one quadrature can be considerably smaller than the 
quantum limit. We found that tunneling effects can enhance the generation of
output oscillator squeezed states.
It was demonstrated that 1D Fokker--Planck equation 
is a quasiclassical limit of a quantum master
equation.  
\end{abstract}

\maketitle

\section{Introduction}
Complex systems with two or more stable states appear in many fields of science from
biology and chemistry to quantum optics and electronics \cite{Innes2013} 
\cite{Semenov2016}
\cite{Hoang2018} \cite{Ma2018}. 

Ability to control and manipulate 
these complex systems 
relies on one's knowledge of 
their stable states, the extent of their robustness with 
respect to environmental fluctuations, and on ability to 
make specific perturbations inducing transitions between 
these states.


First, one needs to understand the behavior of bistable systems interacting
with the environment. Bistable systems in optics and electronics are 
widely used as switching elements in communications systems, 
basic elements of memory devices,
logic gates, optical turnstiles, etc. So, the investigation of fluctuation--induced
transitions between different stable states
is crucial to improve the stability of optical and electronic devices
and to control their switching rates.

Due to unprecedented miniaturization of optical and electronics devices 
quantum effects became very important for their operation 
\cite{Kasprzak2010} \cite{Albert2013}.
It is impossible to study bistability 
without accounting for quantum effects. Thus, it is necessary to trace the correspondence
between the classical and quantum regimes of the system \cite{Kasprzak2013}. 
In the quantum regime it is important
to understand whether quantum fluctuations impose a fundamental limit on stability of optical
and electronic devices.

It is well established that in many nonlinear optical and electronic interface systems 
there exist a set of quantum states
--- squeezed states --- which have less uncertainty in one quadrature than a coherent 
state \cite{Walls1983}.
Generation of squeezed states is a key for implementation of quantum
information protocols and for stability enhancement of quantum 
optics devices \cite{Savage1986}.  
Bistable quantum optics
systems are promising candidates to realize the squeezed states. 
Recently the squeezed exciton--polariton field has been observed 
in pillar--shaped semiconductor microcavities in the bistable regime near the critical point
of the bistable curve \cite{Boulier2014}.

A driven nonlinear oscillator interacting with a thermal bath is the minimal model describing
fluctuation--induced transitions in bistable systems out of equilibrium.
The dynamics of various microcavities coupled with nonlinear media and coherently driven by 
an external field including exciton--polaritons in semiconductor microresonators with 
external pumping can exhibit a bistable behavior and can be described by the model of a 
driven nonlinear oscillator. 
Recent experiments demonstrated that 
as external coherent pumping is increased 
the occupied exciton--polariton mode shows strong sudden jumps 
from one state to another. Such behavior is caused by the fluctuation--induced transitions
between the stationary states. These transitions could also lead to decrease of 
the hysteresis area of an internal microcavity field under the S--shaped response curve 
with respect to the external pumping \cite{Rodriguez2017}.

Another experimental realization which can be analyzed using the nonlinear oscillator 
model is a mesoscopic Josephson junction array resonator \cite{Muppalla2018}.
In such a device, the anharmonicity
can be of the same order as the linewidth, and the dynamics of bistability 
has been experimentally measured by observing the jumps between different stable states.
It was shown experimentally that the switching rate strongly depends on the 
pumping intensity.

In addition, the model of a driven nonlinear oscillator
is applicable to highly excited molecular vibration modes in the presence of
anharmonicity.

The model of a driven nonlinear oscillator has been extensively theoretically studied
since 1980's.
However, fluctuation--induced transitions between two stable states were traditionally
analyzed using the classical 1D Fokker--Planck equation (FPE) 
in quasienergy space without accounting
for quantum tunneling \cite{Vogel1988}, \cite{Maslova1986}, \cite{Dmitriev1986}. 

The ultra--quantum limit of dispersive bistability was analyzed
by Drummond and Walls \cite{Drummond1980}, where the kinetic equation for generalized 
Glauber function was solved analytically for the case of zero bath temperature.  
The same model was analyzed numerically using the technique of quantum master 
equation \cite{Risken1987}. 

Nevertheless, there has been no detailed analysis 
of kinetics of the nonlinear driven oscillator
allowing one to trace the transition between the classical and quantum descriptions of this 
system.
Moreover, the structure of quasienergy states of a driven quantum nonlinear oscillator
and the influence of their degeneracy \cite{Dykman2005} 
on kinetics still remains a relatively unexplored area of research. In order 
to understand rich physical properties of bistable systems 
one could start by considering the minimal model of a driven nonlinear oscillator.

In present work, we derive the quasiclassical kinetic equations taking into account the tunneling 
effects. These equations are a quasiclassical limit of the quantum master equation
for the density matrix of a quantum driven nonlinear oscillator. We show that in the 
quasiclassical limit, tunneling transitions reduce the threshold value of intensity of the
external field responsible for switching between the most probable states
of the system. We also show that tunneling between trajectories in different regions of 
the phase space and multi--photon resonant transitions 
between the states of the nonlinear oscillator
are the same effects.
In the quantum case, we explore the structure of eigenstates and show that
the quasienergy states corresponding to the higher amplitude stable state are squeezed, and
the uncertainty in one of the quadratures can be much lower than the usual quantum limit.

\section{Classical bistability}
\subsection{The basic model}
We consider a model system consisting of a single oscillator mode with Kerr--like 
nonlinearity excited by a resonant field. Its key feature is the bistability
in a certain range of external pumping intensity:
the presence
of two different classical stable states.

The effective Hamiltonian for such a model is \cite{Maslova1986}, \cite{Vogel1990},
\cite{Maslova2007}
\begin{equation}
    \label{rot_wave_ham}
    H_\mathrm{eff} =  -\Delta |a|^2 + \frac{\alpha}{2}|a|^4 - f(a + a^*),
\end{equation}
where $a, a^*$ are slowly varying amplitudes of the internal oscillator field;
$\Delta \equiv \Omega - \omega_0$ is the 
detuning between the external field frequency $\Omega$ and the frequency of the
resonance $\omega_0$; $\alpha$ is the anharmonicity parameter; 
$f$ is the interaction strength with external field (proportional to its
amplitude).
Such a model can arise for various systems in the rotating--wave approximation
such as microcavity with a nonlinear medium coherently driven
by an external field. For example, this effective Hamiltonian can be derived for 
the Janes--Cunnings model after adiabatically excluding the atomic variables. 
It also describes 
the microcavity exciton--polaritons driven by an external field
as well as strongly excited vibration modes of 
molecules in the presence of an external resonant field.
Here we use the normalized field amplitude,
$A \equiv a\sqrt{\alpha/\Delta}$, and a dimensionless time, $\tau = \Delta t$.
The only dimensionless parameter which governs the system dynamics is 
$\beta \equiv \alpha f^2/\Delta^3$.  
\textcolor{black}{
In dimensionless variables the parameter $\beta$ can be treated as the rephasing rate of 
the nonlinear driven oscillator \cite{Lamb1972}. This parameter can also be identified with
the Dicke cooperation parameter determining the typical rate of the intensity growth of a
superradiance pulse. Note that the original Dicke model 
deals with collective superradiance of the system of
quantum two--level emitters interacting with the cavity field. However, as it was shown in 
Refs. \cite{Maslova1988} and \cite{Kocharovsky2017}, a superradiance pulse 
can also arise in a classical system of nonlinear 
oscillators coupled to the cavity field due to rephasing processes.}

In terms of new variables the dimensionless Hamiltonian is given by 
$\mathcal{H} = (\alpha/\Delta^2) H_\mathrm{eff}$:
\begin{equation}
    \label{dimless_ham}
    \mathcal{H} = -|A|^2 + \frac{1}{2}|A|^4 + \sqrt{\beta}(A + A^*),
\end{equation}
while the equation of motion reads
\begin{equation}
    \label{equation_of_motion}
    i\frac{\p A}{\p \tau} = -A + A|A|^2 + \sqrt{\beta}.
\end{equation}

\begin{figure}[h]
    \centering
    \includegraphics[width=\linewidth]{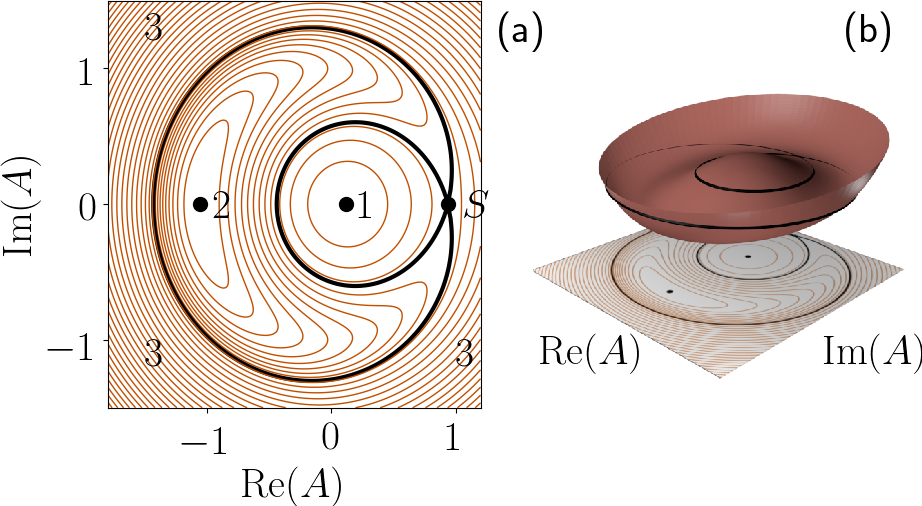}
    \caption{(a) The contour lines of the classical
            Hamiltonian \eqref{dimless_ham} for 
            $\sqrt{\beta/\beta_\mathrm{crit}} = 0.3$. The separatrix (black solid line) 
            divides the plane into three regions: region 1 containing the stationary value
            with a lower amplitude, region 2 containing the stationary value
            with a higher amplitude, and the outer region 3. The unstable stationary
            state S is the point of self--intersection of the separatrix. 
            (b) The 
            same set of contour lines is shown together with the surface plot of the 
            Hamiltonian \eqref{dimless_ham}. 
            It illustrates that the stationary states, $1$, $2$, and
            $S$ correspond respectively to the maximum, minimum, and the saddle point 
            of the Hamiltonian. For each trajectory
            in region $1$, there exists a trajectory in region $3$ with the 
            same quasienergy.}
    \label{fig:phase_portrait}
\end{figure}

The classical phase trajectories of the nonlinear oscillator 
in the plane $(A, A^*)$ are the contour lines of the classical 
Hamiltonian function \eqref{rot_wave_ham} [Fig.~\ref{fig:phase_portrait}(a)].
Let us focus on the structure of \eqref{rot_wave_ham} as the function of two variables,
$\Re A$ and $\Im A$.

At $\beta = 0$, the function has a shape of the Mexican hat potential. It is radially symmetric, 
and its contour lines are concentric circles. 
At nonzero $\beta$, $0 < \beta < \beta_\mathrm{crit} = 4/27$, 
the hat is deformed, as shown in Fig.~\ref{fig:phase_portrait}(b). 
Instead of infinitely many local minima, two extrema arise: a 
true local minimum and a saddle point. 

\begin{figure}[h]
    \includegraphics[width=0.9\linewidth]{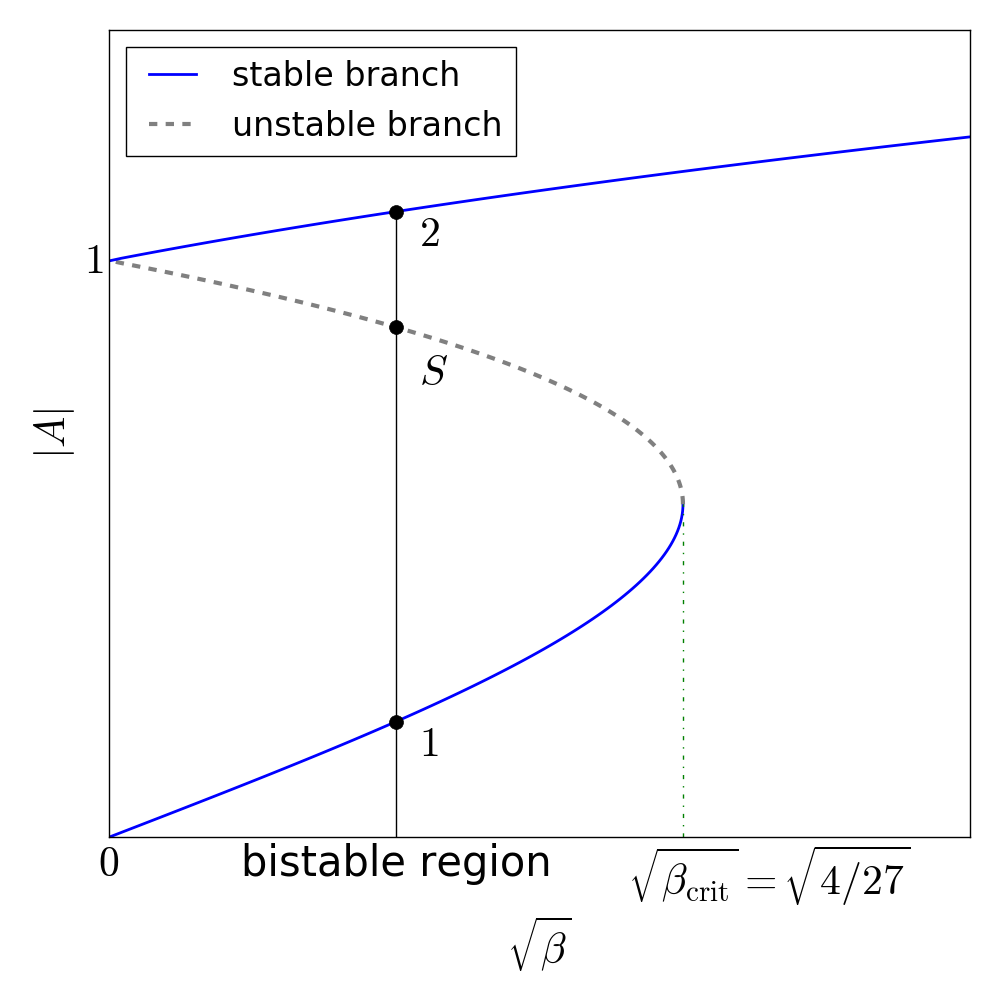}
    \caption{The S--shaped response curve of the normalized amplitude to the external field.
             At $\beta<\beta_\mathrm{crit} \equiv 4/27$
             there are three stationary states. The dashed line corresponds to
             the unstable stationary state.}
    \label{fig:s_shaped_curve}
\end{figure}

The stationary values of $a$ 
are given by the stationary solutions of Eq.~\eqref{equation_of_motion}, 
which defines the S--shaped response curve
(Fig.~\ref{fig:s_shaped_curve}) of the internal field 
amplitude to the external field.

In the bistability region $0 < \beta < \beta_\mathrm{crit}$, 
there are two stable stationary states 1, 2 and one unstable state S,
which lies on a self--intersecting trajectory called separatrix. It divides
the phase plane into three regions: the two inner regions 1 and 2 with the 
corresponding stable states inside them and the outer region 3.
The stable 
state in region $1$ has a lower field amplitude, while
the stable state in region $2$ has a higher field amplitude.

\subsection{Fokker--Planck equation in the presence of white noise}
In any realistic system, noise and damping due to interactions with the environment
are always present. They result in appearance of the damping term with dimensionless 
damping constant $\vartheta \equiv \gamma/\Delta$ and additional random field $\xi$
in the right--hand side of the equations of motion.
\begin{equation}
    \label{equation_of_motion_w_noise}
    \begin{gathered}
       i\frac{\p A}{\p \tau} = -A(1 + i\vartheta) + A|A|^2 + \sqrt{\beta} + \xi,\\
       \avg{\xi(\tau)\xi^*(\tau')} = Q\delta(\tau-\tau'), \quad
       \vartheta \equiv \frac{\gamma}{\Delta}.
    \end{gathered}
\end{equation}
The effect of damping is that the field amplitude 
relaxes to one of the stable stationary states.
Noise has the opposite effect. First, it results in small 
random deviations from the stationary states.
Second, it can induce transitions between the 
stationary states. At weak noise intensity,
these transitions are exponentially rare.

In the case of the white noise \eqref{equation_of_motion_w_noise}, 
it is possible to derive the FPE for the probability density \cite{Johne2009}. 
\begin{equation}
    \label{full_fokker_planck}
    \frac{\p \mathcal{P}}{\p t} = 
    \frac{\p}{\p A} \left(i\mathcal{P}\frac{\p H}{\p A^*} + \vartheta A \mathcal{P} + 
                        \frac{Q}{2}\frac{\p \mathcal{P}}{\p A^*}\right)
    + \mathrm{c. c.}
\end{equation}
    Since the transitions between the stationary states are very rare, the relaxation 
    consists of two stages. At first, the relaxation to the quasi--stationary
    distribution occurs in each region of the phase space at time scales 
    determined by the inverse damping constant.
    Then, at a much slower rate, the 
    probability distribution evolves to the true stationary distribution 
    due to noise--induced transitions between the stable states.

At small damping ($\vartheta \ll 1$) and weak noise ($\vartheta/Q \gg 1$),
a significant simplification of the 2D FPE is possible. Weak damping
and noise give only a small correction to the motion along the phase trajectories.
So, it is natural to average the distribution function in each region of the phase space
along the trajectory and define the approximate
$P_i(t, H(A,A^*))$, $i=1,2,3$. 

Different trajectories with the same quasienergies can exist in regions 1 and 3 
[Fig.~\ref{fig:phase_portrait}].
By averaging the full FPE, one gets the 1D FPE 
in quasienergy space \cite{Vogel1990}, \cite{Dmitriev1986}:
\begin{equation}
    \label{energy_fokker_planck}
    \begin{gathered}
        \frac{\p P_i}{\p t} = \frac{1}{T_i(E)} \frac{\p J_i}{\p E}, \\
        J_i(E) = \vartheta K_i(E) P_i + 
            Q D_i(E) \frac{\p P_i}{\p E}.
    \end{gathered}
\end{equation}
The expressions for $K_i(E), D_i(E), T_i(E)$ were derived in Refs.\cite{Vogel1990}, 
\cite{Maslova1986}, \cite{Dmitriev1986}, \cite{Maslova2007} and are reproduced 
in the Appendix. $T_i(E)$ is the period of motion along the trajectory with 
quasienergy $E$ in the region $i$, 
and $K_i(E)$ and $D_i(E)$ are the drift and the diffusion coefficients in 
quasienergy space in the region $i$.

This Fokker--Planck equation should be solved in every region of the phase space. The full 
solution should be obtained by applying the boundary conditions near the separatrix, which
include the continuity of the probability distribution and the conservation of the flow:
\begin{equation}
    \begin{gathered}
        P_1(E_\mathrm{sep}) = P_2(E_\mathrm{sep}) = P_3(E_\mathrm{sep}),\\
        J_2(E_\mathrm{sep}) = J_1(E_\mathrm{sep}) + J_3(E_\mathrm{sep}).
    \end{gathered}
\end{equation}
The stationary distribution can be obtained by setting the flow $J_i(E)$ to zero, if 
the tunneling effects are neglected (the discussion of the tunneling effects is 
given below).

\subsection{Relative occupation of two stable states}
The general formula for the stationary distribution function follows immediately from 
\eqref{energy_fokker_planck}:
\begin{equation}
    \label{stat_distr}
    P_i^\mathrm{st}(E) = C e^{S_i-S_i(E)},
\end{equation}
\begin{equation}
    \begin{gathered}
        S_i(E) \equiv \frac{\vartheta}{Q} \int_{E_i}^{E} \frac{K_i(E)}{D_i(E)} dE, \quad
        S_i \equiv S_i(E_\mathrm{sep}).
    \end{gathered}
\end{equation}

The distribution has maxima 
in the vicinity of states $1$, $2$, i.e., at the corresponding 
quasienergies $E_1$, $E_2$ \cite{Maslova2007}, \cite{Vogel1990}.
Outside the neighborhood of $E_1$ and $E_2$, $P^\mathrm{st}_i(E)$ is exponentially
small. 
Depending on whether $S_1 > S_2$ or $S_1 < S_2$,
the probability density is mostly concentrated around either 
state $1$ or state $2$. 

Numerical evaluation of $S_1$ and $S_2$ shows that 
$S_1 = S_2$ at $\sqrt{\beta_0/\beta_\mathrm{crit}} = 0.29$. 
Therefore $\beta_0$ corresponds to the threshold pumping intensity:
at $\beta < \beta_0$ the oscillator mostly
remains in state 1 with a small amplitude, and at $\beta > \beta_0$ it mostly remains in
state 2 with a large amplitude.  
Thus the choice of the most probable state is defined by a 
single parameter $\beta$, and the switching from one most probable state to another
occurs at the universal threshold value $\beta = \beta_0$. 
The width of the threshold region is determined by the characteristics of the 
noise. When $|\beta - \beta_0| \sim Q/\vartheta$, both states have
comparable probabilities.

\begin{figure}[h]
    \centering
    \includegraphics[width=\linewidth]{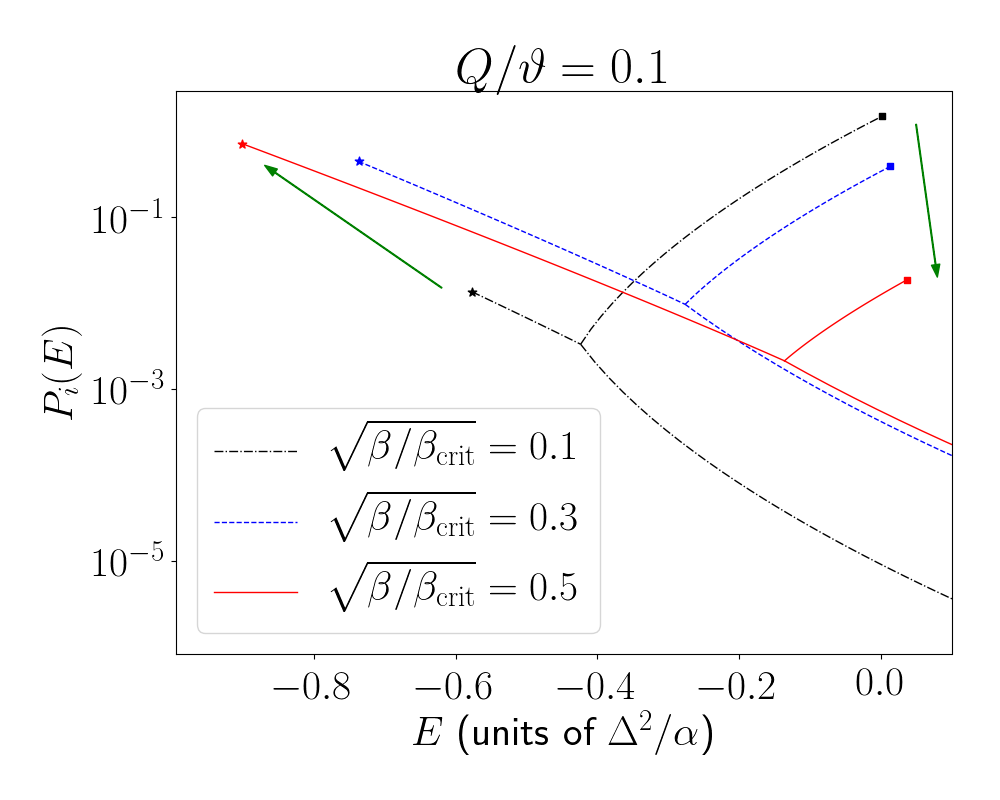}
    \caption{Stationary distribution functions of a classical nonlinear oscillator
             are shown for different values of $\beta$ and $Q/\vartheta = 0.1$.
             Each distribution is single--valued at
             $E_2(\beta) < E(\beta) < E_\mathrm{sep}(\beta)$ and $E > E_1(\beta)$
             and double--valued at $E_\mathrm{sep}(\beta) < E(\beta) < E_1(\beta)$,
             because in the latter case, there exist trajectories
             from regions $1$ and $3$ with the same quasienergy. The distributions have maxima
             at $E = E_{1,2}(\beta)$, which are shown by squares ($E_1$) and stars ($E_2$).
             The arrows show how the distribution function changes with increasing $\beta$.
             At $\beta \approx \beta_0$, the maxima have
             the same order, and at $\beta < \beta_0$ ($\beta > \beta_0$), 
             the maximum at $E_1$($E_2$) dominates.}
    \label{y_shaped_curves}
\end{figure}

\subsection{Transition rates between different stable states}
The relaxation of a nonlinear driven oscillator happens in two stages. 
The first stage is the fast relaxation
to the quasi--stationary distribution which occurs independently in regions $1$ and $2$. 
After that, the slow 
relaxation to the real stationary state occurs, which is governed by 
rare fluctuation--induced transitions 
between the stable states.

\begin{figure}
    \includegraphics[width=\linewidth]{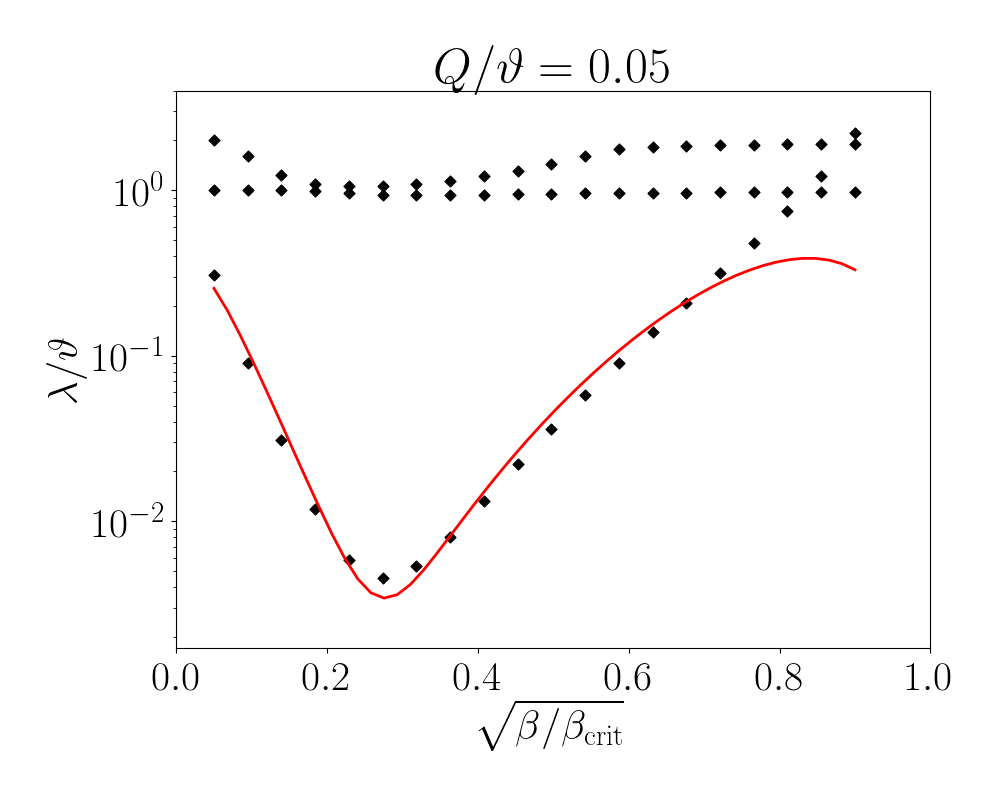}
    \caption{The exact nonzero eigenvalues of the FPE at 
             $Q/\vartheta = 0.05$
             (black diamonds)
             are compared 
             with the asymptotic formula \eqref{relaxation_analytical} (red line). 
             The lowest 
             nonzero eigenvalue is well below $\gamma$ and all other eigenvalues, which
             have the order of $\gamma$. The asymptotic formula fits well the 
             true lowest eigenvalue when it is smaller than $\vartheta$, and is not too close 
             to the edges of the bistability region.}
    \label{fig:class_fp_eigvals}
\end{figure}

Every solution of the FPE can be expressed as a sum over 
eigenfunctions:
\begin{equation}
    P_i(t, E) = \sum_\lambda P^{\lambda}_i(E) e^{-\lambda t},
\end{equation}
where $\lambda$ and $P^{\lambda}$ are the solutions of the eigenvalue problem
\begin{equation}
    -\lambda P^{\lambda}_i = \frac{1}{T_i(E)}
        \frac{\p}{\p E} \left[ \vartheta K_i(E)  + 
        Q D_i(E) \frac{\p}{\p E} \right]P^{\lambda}_i.
\end{equation}
The eigenvalues of the FPE provide an important information about the
kinetics of the system. As shown in Fig.~\ref{fig:class_fp_eigvals}, in the bistability region 
the lowest nonzero eigenvalue is several orders of magnitude smaller than the rest of 
eigenvalues. It determines the last stage of the relaxation process which was described above.

At small $Q/\vartheta$, the lowest eigenvalue $\lambda$ is exponentially small. 
Thus it is possible to use the perturbation theory for $P^\lambda_i$ \cite{Vogel1990}. 
In each region of the phase space, the distribution function up to the first order in 
$\lambda$ is given by
\begin{equation}
    \label{perturbative_solution}
    \begin{gathered}
        P^\lambda_i(E) = P^\mathrm{st}_i(E)\left[1 + \frac{1}{Q}
        \int_{E_\mathrm{sep}}^{E} \frac{\Phi_i(E')dE'}{D_i(E')P^\mathrm{st}_i(E')}\right], \\
        \Phi_i(E) = -\lambda \int_{E_i}^{E} dE'T_i(E')P^\mathrm{st}(E'),
    \end{gathered}
\end{equation}
where $P^\mathrm{st}(E)$ is the stationary distribution \eqref{stat_distr}.
Using the continuity of the probability distribution and the conservation of the flow, 
one gets the following expression for the lowest eigenvalue $\lambda$:
\begin{equation}
    \label{relaxation_analytical}
    \lambda = \frac{\vartheta^2}{Q}\cdot
        \frac{K_1(E_\mathrm{sep})K_2(E_\mathrm{sep})}{K_2(E_\mathrm{sep})-K_1(E_\mathrm{sep})}
            \left[\frac{e^{-S_2}}{D_2'(E_2)} - 
                  \frac{e^{-S_1}}{D_1'(E_1)}\right].
\end{equation}
It is clear that the analytical expression \eqref{relaxation_analytical}
fits well the numerical results everywhere in the bistability region except in the vicinity
of its edges [Fig.~\ref{fig:class_fp_eigvals}].

The lowest eigenvalue non--monotonically depends on the value of $\beta$ and achieves 
its minimum at $\beta = \beta_0$. At $\beta < \beta_0$ ($\beta > \beta_0$),
it corresponds to the escape rate $\lambda_{21}$ ($\lambda_{12}$)
from the higher (lower) amplitude state to the lower (higher) one, which drops (rises) 
with the growing external field intensity.
At the threshold intensity $\beta_0$, $\lambda_{12}$ and
$\lambda_{21}$ have the same values.

\subsection{A tunneling term in the Fokker--Planck equation}
\label{sec:tunnel_fokker_planck}
The trajectories in regions $1$ and $3$ can have the same quasienergy. Thus there
is a possibility of quantum tunneling between them. In the quasiclassical language, it 
can be described as the tunneling term in the FPE:
\begin{equation}
    \label{fokker_planck_w_tunnel}
    \begin{gathered}
        \frac{\p P_{(1,3)}}{\p t} = \frac{1}{T(E)}\frac{\p J_{(1,3)}}{\p E}
            + \lambda_T(E)(P_{(3,1)} - P_{(1,3)}).\\
    \end{gathered}
\end{equation}

Here, $\lambda_T$ is the tunneling rate. It can be calculated in the quasiclassical 
limit with the tunneling amplitude proportional to
$e^{-\frac{\Delta}{\hbar\alpha}\Stunn(E)}$ using the Fermi's golden rule: 
\begin{equation}
    \lambda_T/\vartheta \sim e^{-\frac{2\Delta}{\hbar\alpha}\Stunn(E)}.
\end{equation}

At small $\lambda_T$, the stationary solution $P^T_i(E)$
can be obtained by the perturbation approach similar to 
\eqref{perturbative_solution}:

\begin{equation}
    \begin{gathered}
        P^T_{(1,3)}(E) = P^\mathrm{st}_{(1,3)}(E)\left[1 + \frac{1}{Q}
                \int_{E_\mathrm{sep}}^{E} 
                \frac{\tilde{\Phi}_{(1,3)}(E')dE'}{D(E')P^\mathrm{st}_{(1,3)}}\right], \\
        \tilde{\Phi}_{(1,3)}(E') = \int_{E_{(1,\mathrm{sep})}}^{E'}dE''T(E'') 
           \lambda_T(E'')(P_{(1,3)}^\mathrm{st} - P_{(3,1)}^\mathrm{st}).
    \end{gathered}
\end{equation}
From this equation, we can determine the most probable quasienergy states in regions 
1 and 3 from which the
tunneling occurs. It is defined by the minimum of $S_\mathrm{total}$, 
\begin{equation}
    S_\mathrm{total} = \frac{2\Delta}{\hbar\alpha}\Stunn(E) + \frac{\vartheta}{Q}S_i(E).
    \label{Stotal}
\end{equation}

For calculation of the quasiclassical tunneling exponent $\Stunn$ it is necessary to rewrite
the classical Hamiltonian \eqref{rot_wave_ham} in real variables $p$ and $q$: 
$A = \frac{q + ip}{\sqrt{2}}$:
\begin{equation}
    \mathcal{H} = -\left(\frac{q^2+p^2}{2}\right) + 
        \frac{1}{2}\left(\frac{q^2+p^2}{2}\right)^2 - \sqrt{2\beta}q.
\end{equation}
The tunneling action is defined as an integral over the classically inaccessible area 
$p^2 < 0$:
\begin{equation}
    \Stunn = \int_{q_1}^{q_c} |p_1|dq + \int_{q_c}^{q_2} |p_3| dq.
\end{equation}
The functions $p_{1,3}(q)$ for a specific quasienergy $E$ are determined as 
the solutions of the equation $\mathcal{H}(p,q) = E$:
\begin{equation}
    p^2_{1,3} = 2 - q^2 \pm 2\sqrt{1 + 2E - 2\sqrt{2\beta}q}.
\end{equation}
The turning points $q_1$, $q_2, q_c$ are defined by the
conditions $p_1^2(q_1) = 0$, $p_3^2(q_2) = 0$, 
and $p_1^2(q_c) = p_3^2(q_c)$.

The resulting quasiclassical tunneling exponent has an integral representation
\begin{equation}
    \Stunn = \int_{q_1}^{q_2} \acosh\left\{ 
        \frac{E + \frac{x^2}{2} - \frac{x^4}{8}}{\sqrt{2\beta}x}\right\} x\,dx.
\end{equation}

At small $\beta/\beta_\mathrm{crit}$, it can be approximated as 
\begin{equation}
    \Stunn = \sqrt{1 + 2E}\ln{\frac{1}{\beta}} + O(1).
\end{equation}
Now, from the expression \eqref{Stotal} we can estimate the quasienergy state which is optimal
for tunneling.

At $\vartheta/Q \gg \Delta/\hbar\alpha$,
$S_\mathrm{total}$ has a minimum near $E \sim E_1$, 
Therefore tunneling transitions occurs directly between
the lower--amplitude stable state and the corresponding state from region 3. 
On the contrary, at $\vartheta/Q \ll \Delta/\hbar\alpha$ 
the ''total action''
$S_\mathrm{total}$ has a minimum near $E \sim E_\mathrm{sep}$.
So, tunneling occurs between the states with quasienergy close to $E_\mathrm{sep}$,
and the noise--induced transitions dominate.

We concentrate on the case $\vartheta/Q \gg \Delta/\hbar\alpha$. In this limit, 
the leading term in the tunneling action at $E = E_1$ is 
\begin{equation}
    \Stunn(E_1) = \ln{\frac{1}{\beta}} + O(1).
\end{equation}

The pre--exponential factor has the order of $\Delta$. It can be evaluated by 
matching the quasiclassical solutions near the turning points.

Tunneling between the quasiclassical trajectories effectively occurs when
the quasienergies obtained from the Bohr--Sommerfeld quantization rule become almost equal.
In this case, tunneling leads to an exponentially small splitting between them. As will
be shown below, even at finite $\beta$, this occurs when
$2\Delta/\hbar\alpha$ is exactly integer. 
In this case, the tunneling rate between the 
classical trajectories in regions 1 and 3 with closest quasienergies is estimated as
\begin{equation}
    \label{tunn_prob_1}
    \lambda_T \propto \frac{\Delta^2}{\gamma} \beta^{\frac{2\Delta}{\hbar\alpha}}.
\end{equation}
When $2\Delta/\hbar\alpha$ is not integer, and $\gamma \ll \Delta$, one has
\begin{equation}
    \label{tunn_prob_2}
    \lambda_T \propto \gamma \beta^{\frac{2\Delta}{\hbar\alpha}}.
\end{equation}

In both cases, the tunneling rate is proportional to $\beta^{\frac{2\Delta}{\hbar\alpha}}$. 
At integer $2\Delta/\hbar\alpha = m$, the tunneling rate
can be treated as the probability of $m$--photon resonant
transition between the real energy states of the nonlinear oscillator. So, the tunneling processes
in the presence of a resonant external field and the multi--photon transitions between
the energy states of a nonlinear oscillator are the similar 
effects \cite{Keldysh1965}.

The same expression for the tunneling amplitude in the lowest non--vanishing order 
can be also obtained in the framework of the
quantum--mechanical perturbation theory for multi--photon transitions \cite{Larsen1976}:
\begin{equation}
    \label{splitting}
    A_{k,m-k} = \frac{\Delta}{m} \left(\frac{\beta m^3}{2}\right)^{\frac{m}{2} - k}
                \frac{\sqrt{(m-k)!}}{(m-2k-1)!^2\sqrt{k!}}.
\end{equation}
For $k = 0$,
\begin{equation}
    |A_{0,m}|^2 \propto \Delta^2 \beta^m.
\end{equation}
The state with $k = 0$ corresponds to the point 1 on the phase portrait. 
So, for a driven bistable system, the probability of $m$--photon transition 
calculated quantum--mechanically \eqref{splitting} is the same as the
tunneling probability between the degenerate quasienergy states in the
quasiclassical treatment.

The same nature of tunneling effects and multi--photon ionization of atoms in a strong
electromagnetic field was first demonstrated by L. V. Keldysh \cite{Keldysh1965}.

The presence of tunneling modifies both the distribution function and the relaxation rate. If
$\lambda_T$ is small, its effect can be taken into account within the perturbation theory.
The ratio of probability densities of states $1$, $2$ modifies as follows:
\begin{equation}
    \label{occupation_ratio_w_tunnel}
    \frac{P_2^\mathrm{st}(E_2)}{P_1^\mathrm{st}(E_1)} = e^{S_2}\left(e^{-S_1} + 
        \frac{\lambda_T(E_1)Q}{\vartheta^2} \frac{D'_1(E_1)}{K_1(E_\mathrm{sep})}\right).
\end{equation}

\begin{figure}[h!]
    \includegraphics[width=\linewidth]{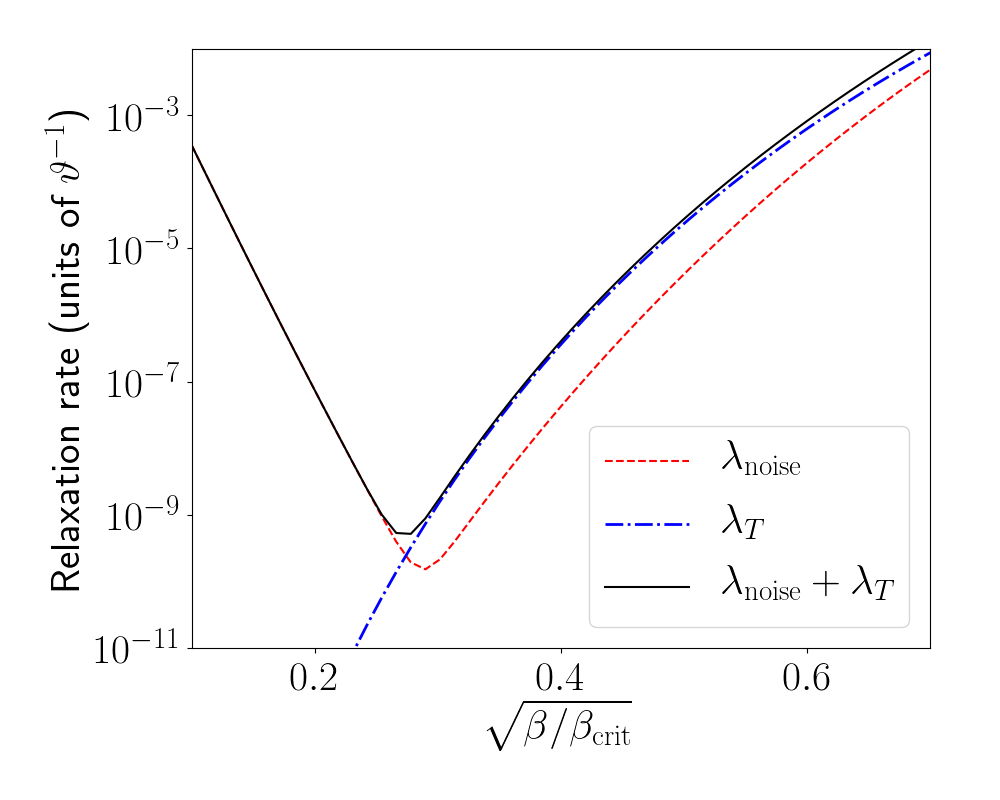}
    \caption{The relaxation rate in units of $\vartheta$ without and with the tunneling term.
             Here $Q = 0.015$ and $\hbar\alpha/\Delta = 0.2$ }
    \label{fig:trans_rate_w_tunnel}
\end{figure}

According to this formula, tunneling leads to a 
decreasing probability to be in state $1$. 
Tunneling also changes the total transition rate between the stable states:
\begin{equation}
    \lambda_\mathrm{total} = \lambda_\mathrm{noise} + \lambda_T(E_1).
\end{equation}
Here, $\lambda_\mathrm{noise}$ is defined by \eqref{relaxation_analytical}. 

The behavior of the transition rate between the stable states in the presence of tunneling
is depicted in Fig.~\ref{fig:trans_rate_w_tunnel}. 
Tunneling transitions shift the threshold value of the 
external field intensity towards lower values and increase the threshold
values of the transition
rate.

\section{Bistability in quantum oscillator}
\subsection{Quantum quasienergy states and squeezing}
The Hamiltonian for a quantum bistable oscillator in the rotating--wave approximation
is given by:
\begin{equation}
    \begin{gathered}
        \label{quantum_bist_ham}
        \hat{H}_0 = -\Delta \ha^\dagger \ha + \frac{\alpha}{2} \ha^\dagger \ha^\dagger \ha\ha 
                + f(\ha + \ha^\dagger),\\
        [\ha,\ha^\dagger]=1.
    \end{gathered}
\end{equation}
The operators $\ha$ and $\ha^\dagger$ are the creation and annihilation
operators of the internal oscillator field. 
In the quasiclassical limit, $\sqrt{\hbar} \ha$ and $\sqrt{\hbar} \ha^\dagger$ correspond to 
the classical field amplitudes. In the following, we set $\hbar = 1$. 

The exact eigenstates of \eqref{quantum_bist_ham} should be obtained numerically by
diagonalization of the Hamiltonian matrix. However, qualitatively, the structure
of eigenstates can be understood using the classical analogy.

From the Bohr--Sommerfeld quantization rule, one concludes that
the eigenstates of the Hamiltonian correspond to the discrete set of 
trajectories in the classical phase portrait (Fig.~\ref{fig:phase_portrait}). 
However, the real picture is a bit more complicated 
because the quantum tunneling should also be taken into account. This is because
the classical phase portrait has different regions with the same quasienergy, i.e., regions 
I and III. So, the real eigenstates may correspond not only to single trajectories but 
also to superpositions of two trajectories with the same quasienergy. 

The possibility of quantum tunneling is closely connected to the degeneracy of eigenstates 
in the Hamiltonian \eqref{quantum_bist_ham} at $f = 0$. 
At $f = 0$, the Hamiltonian commutes with 
$a^\dagger a$, and the states with $k$ excitation quanta 
are the eigenstates of the Hamiltonian.
Their quasienergy is 
\begin{equation}
    \epsilon_k^{(0)} = -\Delta k + \frac{\alpha k(k-1)}{2}.
\end{equation}
For integer $2\Delta/\alpha$, the states with $k$ and $m-k$ excitation quanta become 
degenerate. At small but nonzero $f$, these states can mix: the true eigenstates
are the superpositions of $|k\rangle$ and $|m-k\rangle$. In the quasiclassical language,
this corresponds to tunneling between degenerate classical trajectories. Numerical
diagonalization shows that such mixing occurs only when $2\Delta/\alpha$ is 
very close to an integer. 

To provide some illustration to this qualitative picture, we calculated the eigenstates
of the Hamiltonian \eqref{quantum_bist_ham} in the coherent basis:
\begin{equation}
    \psi_n(z) \equiv \langle n | z \rangle,
\end{equation}
where $|z\rangle$ is a normalized coherent state. The function $\psi_n(z)$ corresponding
to the $n$--th eigenstate has a maximum near the contour line of the classical Hamiltonian
$H(a, a^*) = E_n$. This means that the quantum state $|n\rangle$ corresponds to the 
classical motion along the trajectory $H(a, a^*) = E_n$.
\begin{figure}[h]
    \includegraphics[width=1.0\linewidth]{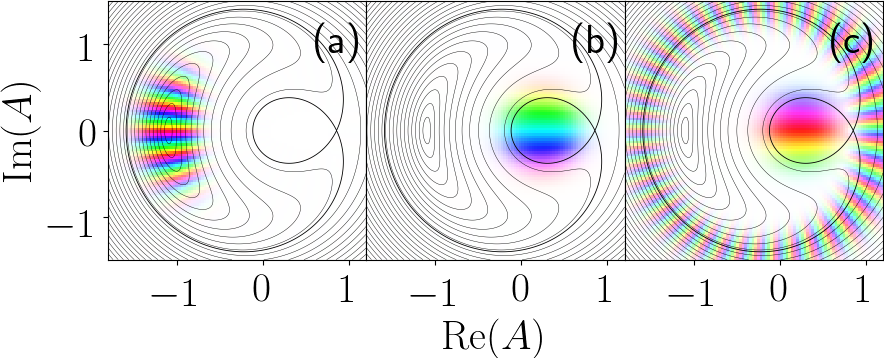}
    \caption{Some eigenstates of the quantum Hamiltonian are shown in the coherent basis.
             For each state $|n\rangle$, we show 
             the quantity $\langle z | n \rangle$ in complex $z$  
             plane ($z = A\sqrt{\Delta/\alpha}$). (a) 
             The higher--amplitude state for $\Delta/\alpha = 20.25$ and
             $\sqrt{\beta/\beta_\mathrm{crit}} = 0.6$. It is squeezed in the
             $q$--direction. (b) The lower--amplitude state for the same parameters.
             (c) The eigenstate which is a superposition of 
             the lower--amplitude state and the trajectory from the classical region $3$.
             It corresponds to $\Delta/\alpha = 20$ and
             $\sqrt{\beta/\beta_\mathrm{crit}} = 0.6$.
             }
    \label{fig:mixed_phase_portrait}
\end{figure}

An important property of the quasienergy states is that 
the states corresponding to the higher amplitude
stable point are squeezed. This can be shown by the mean--field expansion:
\begin{equation}
    \begin{gathered}
        \ha = \avg{a}_2 + \ha'.\\
    \end{gathered}
\end{equation}
The mean value of $\ha$ is defined from the equation 
$\frac{\p H}{\p a}(\avg{a}_2,\avg{a^*}_2) = 0$, which corresponds to the classical stable 
states $i$. For small $\beta$, the mean field value in the higher amplitude stable state 2
is $\langle a \rangle_2 \approx \sqrt{\Delta/\alpha}(1 + \sqrt{\beta}/2)$.

The quadratic part of the Hamiltonian takes the form
\begin{multline}
    \hat{H} = H(\avg{a}_2, \avg{a^*}_2) - 
        (\Delta - 2\alpha|\langle a\rangle_2|^2) \ha'^\dagger \ha'\\
        + \frac{\alpha}{2}\langle a\rangle_2^2 \ha'^\dagger \ha'^\dagger
        + \frac{\alpha}{2}\langle a^\dagger \rangle_2^2 \ha'\ha'.
\end{multline}
We diagonalize this Hamiltonian using the Bogolyubov transformation:
\begin{equation}
    \begin{gathered}
        \ha' = \hat{b}\cosh\theta - \hat{b}^\dagger\sinh\theta,\\
        \tanh{2\theta} = 
        \frac{\alpha|\langle a\rangle_2|^2}{2\alpha |\langle a\rangle_2|^2 - \Delta}.
    \end{gathered}
\end{equation}
Let us consider the uncertainties in two quadratures $\hat{q}$ and $\hat{p}$, 
$\hat{a} = \frac{\hat{q} + i\hat{p}}{\sqrt{2}}$. Squeezing is more pronounced in 
the higher amplitude stable states $\langle a \rangle_2$.
\begin{equation}
    \label{uncertainties}
    \begin{gathered}
        \avg{q^2} - \langle q \rangle_2^2 = 
            \frac{e^{-2\theta}}{2} = 
        \frac{1}{2}\sqrt{\frac{\alpha|\langle a\rangle_2|^2 - \Delta}
                              {3\alpha|\langle a\rangle_2|^2 - \Delta}},\\
        \avg{p^2} - \langle p \rangle_2^2 = 
            \frac{e^{2\theta}}{2} =
        \frac{1}{2}\sqrt{\frac{3\alpha|\langle a\rangle_2|^2 - \Delta}
                              {\alpha|\langle a\rangle_2|^2 - \Delta}}.
    \end{gathered}
\end{equation}
The quadratic approximation is correct 
when $\sqrt{\beta}$ is larger than $\alpha/\Delta$. 
When $\alpha/\Delta \ll 1$,
this is fulfilled almost in the entire region of bistability, and the relations 
\eqref{uncertainties} are valid.

The minimum possible uncertainty of $\hat{q}$ is at 
$\beta \sim (\alpha/\Delta)^2$,
where it can be estimated as
\begin{equation}
    \avg{q^2} - \avg{q}_2^2 \sim \sqrt{\frac{\alpha}{\Delta}}.
\end{equation}
Thus, the uncertainty in $\hat{q}$ quadrature can be far beyond the quantum limit.

As we have shown in the previous section \eqref{occupation_ratio_w_tunnel}, 
the tunneling effects increase the occupation of the stable state 2 with 
a higher amplitude and therefore enhance 
the generation of squeezed states.

\subsection{Quantum kinetic equation}
Let us assume that the system is weakly interacting with the environment:
\begin{equation}
    H_\mathrm{full} = H_0 + \hat{\xi}^\dagger \ha + \hat{\xi} \ha^\dagger + H_\mathrm{bath}.
\end{equation}
We assume that the correlation functions of damping operators are delta--correlated:
\begin{equation}
    \begin{gathered}
        \avg{\hat{\xi}(t)\hat{\xi}^\dagger(t')} = \gamma(N + 1)\delta(t-t'),\\
        \avg{\hat{\xi}^\dagger(t)\hat{\xi}(t')} = \gamma N\delta(t-t'),\\
    \end{gathered}
\end{equation}
where $N$ is the number of noise quanta.

With such assumptions, the density matrix 
evolution can be described by the master equation \cite{Haken1965}, \cite{Risken1965},
\cite{Graham1970}.
\cite{Drummond1980}, \cite{Risken1987}:
\begin{equation}
    \label{lindblad_eq}
    \p_t \rho = i[\rho, H] + \frac{\gamma}{2}
                    (2\ha\rho \ha^\dagger - \rho \ha^\dagger \ha - \ha^\dagger \ha \rho 
                                      + 2N[[\ha,\rho],\ha^\dagger]).
\end{equation}

If $\gamma$ is small compared to $\Delta$, the
density matrix is almost diagonal in the basis of eigenstates $|n\rangle$, 
and the master equation 
reduces to the rate equation for probabilities $P_n$ to be in the $n$--th eigenstate:
\begin{equation}
    \label{rate_eq}
    \begin{gathered}
        \frac{dP_n}{dt} = \sum_{n'} w_{nn'} P_{n'} - w_{n'n} P_n,\\
        w_{nn'} = \gamma
                \left[ 
            (N + 1)| \langle n | \hat{a} | n' \rangle|^2 
           + N |\langle n' | \hat{a}| n \rangle |^2 \right].
    \end{gathered}
\end{equation}
This equation is a quantum analog of \eqref{energy_fokker_planck}.
The evolution of the density matrix has the same features as the evolution of the 
distribution function for a classical oscillator with bistability. 
At infinite time, the density matrix 
evolves to the stationary distribution. The relaxation to $P^\mathrm{st}_n$ consists of two 
stages. The first stage corresponds to the relaxation to the quasi--stationary distribution.
Its typical time is $\gamma^{-1}$. The second stage is the relaxation to the true 
stationary state. This stage is very slow and happens due to transitions 
between the classical stationary states. These transitions 
can be induced by quantum fluctuations as well as by thermal noise.

Formally, the general solution of \eqref{rate_eq} reads
\begin{equation}
    \label{cauchy_problem}
    P_n(t) = P_n^\mathrm{st} + 
            \sum_{\lambda > 0} C_k P^{\lambda}_n e^{-\lambda t},
\end{equation}
The lowest nonzero eigenvalue is much smaller than all other eigenvalues. Therefore,
at large $t$ only the term with the lowest nonzero $\lambda$ should be retained in 
Eq.~\eqref{cauchy_problem}.

The density matrix relaxes to the true stationary distribution 
with the rate $\lambda_\mathrm{min}$, which 
can be interpreted as the rate of fluctuation--induced transitions between 
the stable states.

\subsection{The quasiclassical limit}
One can show that the continuous limit of \eqref{rate_eq} is the
\eqref{energy_fokker_planck}. As it was mentioned above, every eigenstate corresponds to 
a trajectory on the classical phase portrait, and the hybridization of the
trajectories from regions 
$1$ and $3$ can be neglected unless $2\Delta/\alpha$ is very close to an integer. Thus, in
the quasiclassical limit, the distribution function $P_n$ weakly depends on $n$ in each
of the regions of the phase space. Moreover, the transition rates 
$w_{n, n'} \equiv \tw_{\bar{n},k}$, $\bar{n} = \frac{n+n'}{2}$,
$k = n'-n$
decrease fast with an increasing value of $|k|$ and weakly depend on $\bar{n}$, which is close
to $n$.
In this case, 
it is possible to perform a gradient expansion of $P_n$, $w_{nn'}$ in \eqref{rate_eq}:
\begin{equation}
    P_{n+k} = P_n + \frac{\p P_n}{\p n}k + \frac{1}{2}\frac{\p^2 P_n}{\p n^2}k^2 + \dots,
\end{equation}
\begin{equation}
   \label{w_expansion}
   w_{n,n+k} = \tw_{n,k} 
             + \frac{\p \tw_{n,k}}{\p \bar{n}} \left(\frac{k}{2}\right)
     +\frac{1}{2} \frac{\p^2 \tw_{n,k}}{\p \bar{n}^2} \left(\frac{k}{2}\right)^2 + \dots
\end{equation}
In \eqref{w_expansion}, we took into account that 
$w_{n,n+k} = \tw_{n+\frac{k}{2},k} = \tw_{\bar{n},k}$, $\bar{n} = n + \frac{k}{2}$.
Keeping the terms up to the second order in $k$, one obtains 
the differential equation for 
$P_n$:
\begin{equation}
    \frac{\p P_n}{\p t} = \frac{\p}{\p n}\left[ A(n)P_n + B(n)\frac{\p P_n}{\p n} \right],
\end{equation}
where the coefficients $A(n)$ and $B(n)$ are given by the expressions
\eqref{rate_eq} for 
probabilities $w_{nn'}$:
\begin{equation}
    A(n) = -\sum_{k}\tw_{n,k}k =
    \frac{i\gamma \tilde{T}(\epsilon_n)}{2\pi}\avg{n|\ha\,\p_t \ha^\dagger|n},
\end{equation}
\begin{equation}
    \begin{gathered}
        B(n) = \frac{1}{2}\sum_{k}\tw_{n,k}k^2 = 
            \gamma\left(N + \frac{1}{2}\right) 
            \frac{\tilde{T}(\epsilon_n)^2}{4\pi^2}\avg{n|\p_t \ha\p_t \ha^\dagger |n}.
    \end{gathered}
\end{equation}
Here $\tilde{T}(\epsilon_n)$ is the period of the classical motion with quasienergy 
$\epsilon_n$.

In the quasiclassical limit, the averages over the quantum quasienergy states transform to 
time--averages over the classical trajectories. Thus, 
in the quasiclassical limit, $A(n)$ and $B(n)$ are expressed as
line integrals over the classical trajectories:
\begin{equation}
    \begin{gathered}
        A(n) = \frac{i\gamma}{4\pi}\oint_{C(\epsilon_n)} a\,da^* - a^*da,\\
        B(n) = \frac{i\gamma \tilde{T}(\epsilon_n)}{8\pi^2}
            \left(N + \frac{1}{2}\right)
            \oint_{C(\epsilon_n)} \frac{\p H}{\p a} da - \frac{\p H}{\p a^*} da^*.\\
    \end{gathered}
\end{equation}
After a change of variables $\frac{\Delta}{\alpha} T(E)dE = 2\pi dn$, 
$t\Delta = \tau$, and $\gamma/\Delta = \vartheta$
the equation transforms to the classical FPE
\eqref{energy_fokker_planck}. The coefficient $A(n)$ transforms to $\vartheta K(E)$, and 
$B(n)$ transforms to $QT(E)D(E)$, where 
\begin{equation}
    \label{noise}
    Q = \frac{\vartheta\alpha}{\Delta}\left(N + \frac{1}{2}\right),
\end{equation}
$E = \alpha\epsilon/\Delta^2$ is the dimensionless
quasienergy, and $T(E) = \Delta\tilde{T}(E\Delta^2/\alpha)$ is the dimensionless period 
as in \eqref{energy_fokker_planck}.

\subsection{Results and discussion}
Qualitatively, the behavior of $P^\mathrm{st}_n$ in the diagonal approximation 
resembles the behavior of $P^\mathrm{st}_i(E)$ of a classical oscillator, 
as $P^\mathrm{st}_i(E)$ is the classical limit of $P^\mathrm{st}_n$ (here $i$ indicates 
the classical region of the phase space). 
As $P^\mathrm{st}_i(E)$, it consists of 
two sharp peaks which can be attributed to the classical stable states $1$ and $2$. 
Below (above) the threshold value of the external field, the state $1$ ($2$) dominates.

We directly compared the distributions obtained from the classical FPE 
and from the quantum master equation. 
In the classical limit
$P^\mathrm{st}_n$ equals $(2\pi\alpha/\Delta)P_i^\mathrm{st}(E_n)$, 
$E_n = \alpha\epsilon_n/\Delta^2$,
where $P^\mathrm{st}_i(E)$ is 
the classical distribution function for a dimensionless Hamiltonian \eqref{dimless_ham}
with number of noise quanta defined by \eqref{noise}.
The index $i$ corresponding to the classical region of the phase space is uniquely defined
for each eigenstate unless $2\Delta/\alpha$ is an integer. In the latter case, the
classical 
FPE should be derived from the quantum master equation more carefully. 
It can be obtained only after choosing the proper basic quasienergy states. 
One should deal with the quantum states corresponding 
to the trajectories in the regions of phase space 1 and 3, but not with their superposition.

\begin{figure}[h]
    \centering
     \includegraphics[width=\linewidth]{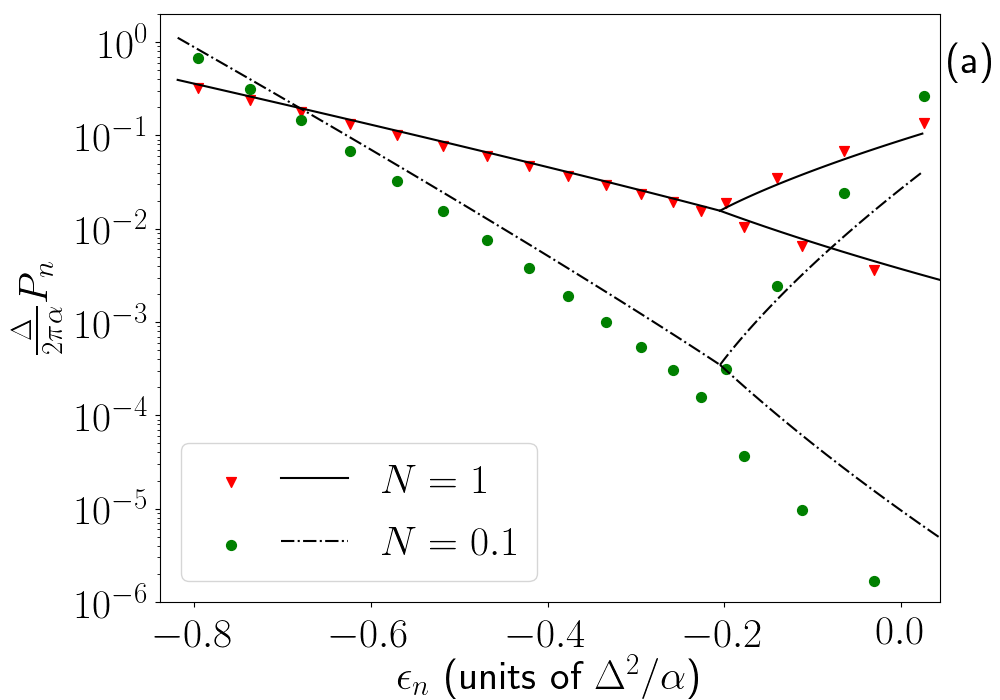}
     \includegraphics[width=\linewidth]{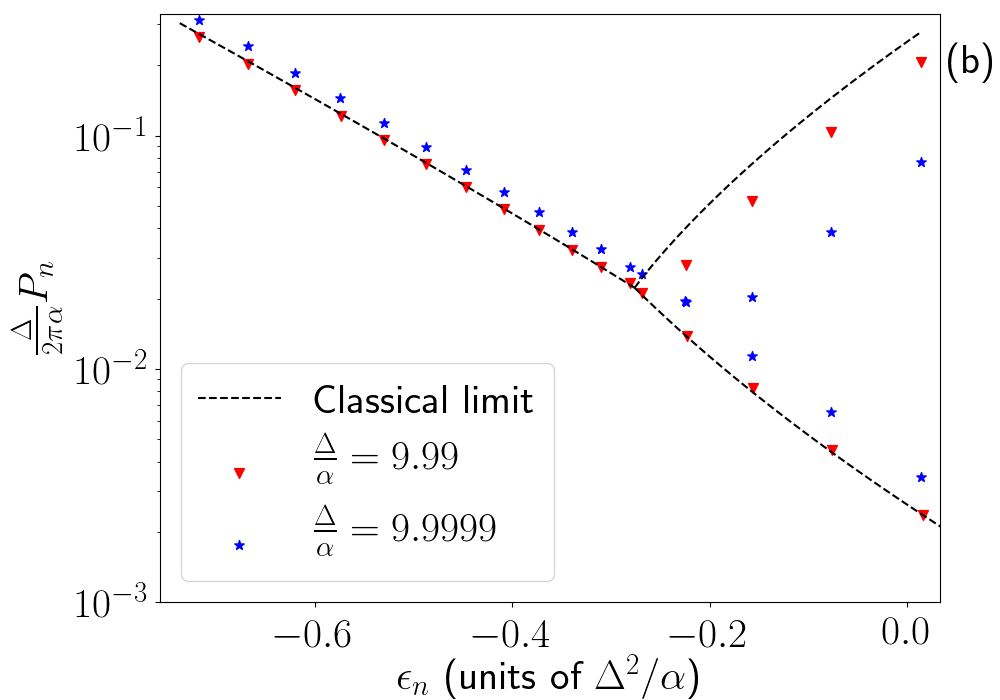}
    \caption{(a) The quantum analog of classical distribution function for 
             $\Delta/\alpha = 9.8$, $\sqrt{\beta/\beta_\mathrm{crit}} = 0.3$
             and $N = 0.1, 1$. Solid (dashed) line represents the
             classical distribution for $N = 1(0.1)$, triangles (circles)
             represent the quantum distribution for $N = 1(0.1)$.
             For $N = 1$, the quantum distribution 
             fits well the corresponding
             classical distribution. For $N = 0.1$, the quantum and
             classical distribution differ significantly. 
             (b) The distribution functions at $N = 1$ and different closely spaced
             values of $\Delta/\alpha$.
             At $\Delta/\alpha = 9.99$ (red triangles),
             the distribution is close to the
             classical distribution (black dashed line). 
             At $\Delta/\alpha = 9.9999$ (blue stars),
             the population in the lower--amplitude state is
             significantly smaller than the prediction of the
             classical FPE.
        }
    \label{fig:energy_distrib}
\end{figure}

For the classical case, it was shown that the change of the most probable stable state 
takes place 
at $\sqrt{\beta} \equiv \sqrt{\alpha f^2/\Delta^3} \approx 0.29$. 
From Fig.~\ref{fig:energy_distrib}, it is clear that for a rather high number of noise quanta 
$N \gg \alpha/\Delta$ and for a large non--integer $2\Delta/\alpha \gg 1$ 
the probability
distribution for a quantum oscillator obtained from the master
equation \eqref{rate_eq} coincides with the classical distribution over quasienergies. For 
a small number of noise quanta, the situation is more complicated 
even for large $\Delta/\alpha$.
Even though at large $\Delta/\alpha$, the quasiclassical approximation for
matrix elements of $\hat{a}$ is valid, the quantum distribution function doesn't coincide 
with the classical one due to quantum fluctuations and tunneling effects.
For a quantum oscillator described by the rate equation 
\eqref{rate_eq}, $\beta_0$ is no more the universal
threshold parameter.
Not only the parameter $\beta$ matters, but also $\Delta/\alpha$ and $N$ matter as well.
The threshold value $\beta_q$ in the quantum limit can exceed the 
classical value for a non--integer $2\Delta/\alpha$, and it is 
below the classical value at integer $2\Delta/\alpha$.
At a small number of noise quanta, when $N \lesssim \alpha/\Delta$, 
the quantum effects including the tunneling processes
govern the transitions between different stable states. In this case,
the distribution function and the relaxation rate strongly depend on 
the value of $\Delta/\alpha$. 

\begin{figure}[h]
    \includegraphics[width=\linewidth]{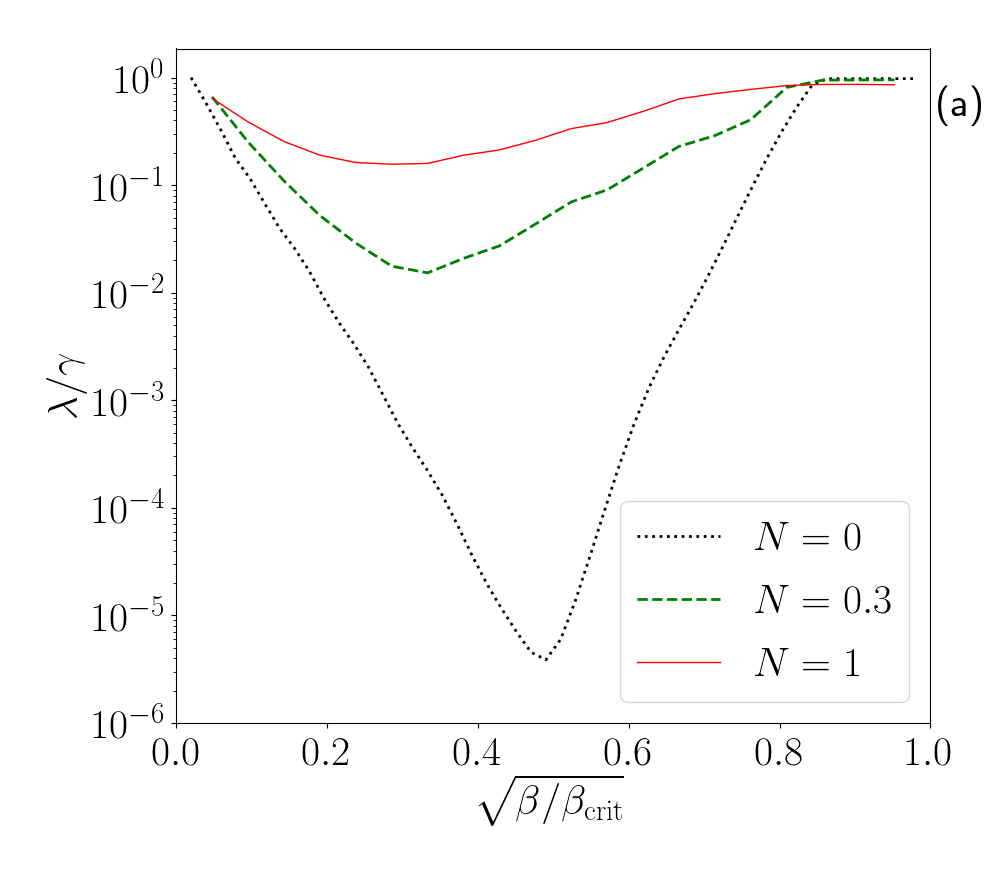}
    \includegraphics[width=\linewidth]{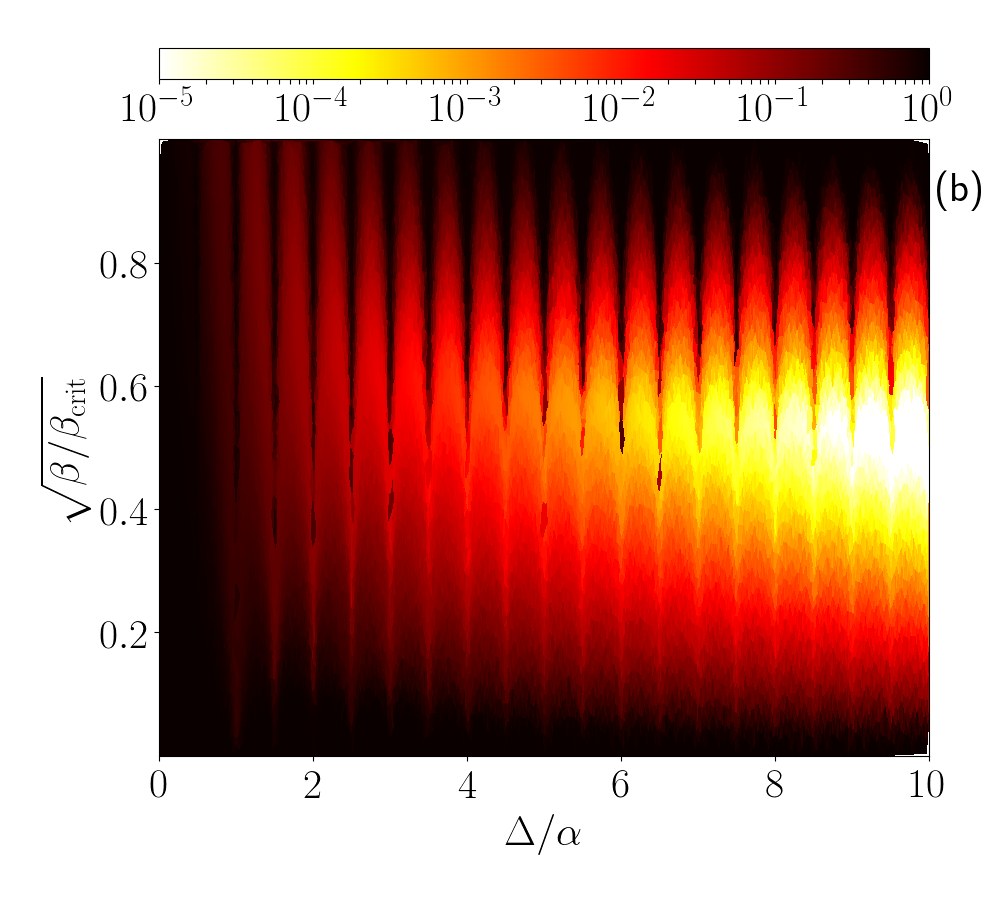}
    \caption{(a) The lowest eigenvalue of the rate equation for $N = 0, 0.3, 1$, and
             $\Delta/\alpha = 9.9$. The black dotted line (green dashed line, red solid line)
             corresponds to $N = 0$ ($N = 0.3$, $N = 1$).
             (b) The lowest eigenvalue
             of the rate equation for $N = 0$ plotted as a function of 
             $\Delta/\alpha$, $\beta$.}
    \label{fig:relaxation_rate}
\end{figure}

When $\Delta/\alpha$ becomes close to an integer, the quantum distribution even at 
$N > \alpha/\Delta$, $\Delta/\alpha \gg 1$ doesn't coincide with
the classical one due to tunneling effects. This is demonstrated in
Figs.~\ref{fig:energy_distrib}(b) and \ref{fig:relaxation_rate}(b). 
Tunneling between the classical regions $1$ and $3$ leads
to a decreasing threshold intensity $\beta_q$ and an increasing relaxation rate, as was mentioned
in Sec.~\ref{sec:tunnel_fokker_planck}.

Moreover, as shown in Fig.~\ref{fig:relaxation_rate}, 
the value of $N$ also influences the threshold external field intensity 
and the threshold relaxation rate.
As $N$ is decreased, the threshold external field intensity rises and reaches
$\sqrt{\beta/\beta_\mathrm{crit}} \approx 0.5$ at $N = 0$. Such behavior 
does not appear in the quasiclassical FPE solutions.

To clarify the influence of the fluctuation--induced transitions on statistical properties
of the internal oscillator field,
one should 
calculate the second--order correlation function $g^2(0)$:
\begin{equation}
    g^2(0) \equiv \frac{\avg{\ha^\dagger \ha^\dagger \ha \ha}}{\avg{\ha^\dagger \ha}^2}.
\end{equation}
When one of the stable states dominates, $g^{(2)}(0) \sim 1$. In a
narrow region of fluctuation--induced transitions $g^{(2)}(0)$ is significantly larger.
For several values of $N$, we have calculated $g^2(0)$ as a function
of two parameters: $\sqrt{\beta}$ and $\Delta/\alpha$. 
\begin{figure}[h]
    \includegraphics[width=0.95\linewidth]{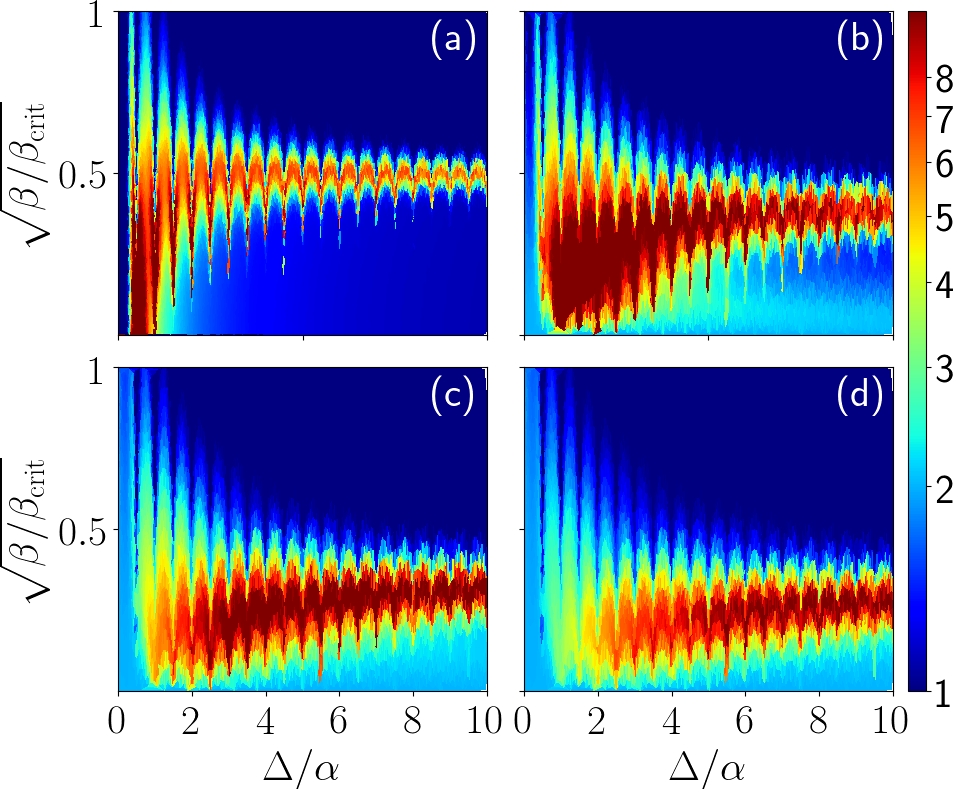}
    \caption{The correlation function $g^2(0)$ at different values of $N$
             plotted as a function of $\Delta/\alpha$, $\beta$. (a) $N = 0$
             (b) $N = 0.01$ (c) $N = 0.05$, and (d) $N = 0.1$.}
\end{figure}

For each $\Delta/\alpha$, the correlation function has similar dependence
on $\sqrt{\beta}$: there is a sharp peak at some value of $\sqrt{\beta/\beta_\mathrm{crit}}$ 
between $0$ and $1$,
which indicates the change of the most probable state. However, the peak is not placed at
$\sqrt{\beta/\beta_\mathrm{crit}} \approx 0.29$ 
as the classical FPE predicts. Its position
is an oscillating function of $\Delta/\alpha$ with sharp minima when 
$2\Delta/\alpha$ is an integer, which corresponds to enhanced tunneling between
degenerate quasienergy states.
The amplitude of oscillations decreases
as $2\Delta/\alpha$ increases, when one approaches the quasiclassical limit. This
fact points to the quantum nature of these oscillations.

\section{Conclusions}
We have derived the quasiclassical kinetic equations for the probability distribution over quasienergy states
of a nonlinear driven oscillator, taking into account the 
tunneling effects. The stationary distribution 
for a wide range of system parameters 
and the typical relaxation rate have been determined. It was shown that
the relaxation 
consists of two stages. At first, the relaxation to the quasi--stationary
distribution occurs in each region of the phase space at time scales 
determined by the inverse damping constant. Then, at exponentially large times, the 
probability distribution evolves to the true stationary one. The relaxation to
the true stationary state happens due to fluctuation--induced
transitions between the quasiclassical stable states.
In the classical limit, if tunneling is neglected, there exists a universal threshold 
value of the field intensity responsible for switching between the most probable stable 
states of the system. Taking into account the tunneling effects renders this value
non--universal. The tunneling transitions lead to a decrease of the threshold value of
external field intensity and to an up to one order of magnitude increase of 
the fluctuation--induced transition rate between the stable states in the threshold area.

For a driven quantum nonlinear oscillator, we demonstrated that the 
quasienergy state corresponding to the classical higher  amplitude
stable state is squeezed. The degree of squeezing is determined by the ratio of nonlinearity
and detuning, and the uncertainty of one of the oscillator quadratures can be much lower 
than the usual quantum limit. 

As tunneling transitions increase the occupation of the 
higher  amplitude stable state, the generation
of squeezed states can be enhanced in the presence 
of tunneling effects.

Also we demonstrated that the quasienergy states become superpositions of trajectories from 
different regions of the phase space. This happens whenever the detuning 
is an integer or half--integer
multiple of the nonlinear shift per quantum. This happens due to a multi--photon resonance 
between the real eigenstates of the nonlinear oscillator. It was shown that such
resonance can be described in terms of tunneling between the quasienergy states in 
different regions of the classical phase space.

The kinetics of the quantum oscillator was investigated using the quantum master equation.
It was shown that in the limit of large detuning--nonlinearity ratio, large number of thermal
photons and in the absence of multi--quantum resonance,
the classical FPE in the quasienergy space is a continuous limit
of the quantum master equation. Importantly, a large value of the
detuning--nonlinearity ratio is not sufficient for the validity of the classical FPE, 
because at a weak noise the quantum effects become 
especially pronounced. 
The relaxation rate and the threshold intensity of the external field are both
very sensitive to the detuning--nonlinearity ratio. At an integer or half--integer 
detuning--nonlinearity ratio, 
the relaxation rate can increase up to several orders of magnitude and the threshold 
value of the external field intensity shifts towards lower values.
In this case, tunneling between degenerate quasienergy states and the multi--photon resonant
transitions between the original states of the nonlinear oscillator can be treated as
the same effect.

Finally, 
it was shown that the second--order correlation function of the internal 
field strongly rises near the threshold pumping intensity, 
which indicates super--Poissonian statistics
of the internal oscillator field.

\begin{acknowledgements}
    This work was supported by RFBR grants 19--02--000--87a and 18--29--20032mk.
\end{acknowledgements}

\bibliographystyle{apsrev4-1}
\input{arxiv_submission_190808.bbl}

\appendix*

\section{The coefficients of the classical Fokker--Planck equation}
The coefficients of the classical FPE 
are defined as follows:
\begin{equation}
    \begin{gathered}
        K(E) = \frac{i}{2}\oint a\,da^* - a^*\,da,\\
        D(E) = \frac{i}{2}\oint \frac{\p H}{\p a} da - \frac{\p H}{\p a^*} da^*,\\
        T(E) = \int da^*da\,  \delta(E - H(a^*, a)).\\
    \end{gathered}
\end{equation}
For them, we obtained the following integral representations:
\begin{equation}
    T(E) = \int \frac{dt}{\sqrt{2f^2t-\left(E+\frac{t}{2}-\frac{t^2}{8}\right)^2}},
\end{equation}
\begin{equation}
    K_i(E) = \int \frac{3t^2/16 - t/4 + E/2}
            {\sqrt{2f^2t-\left(E+\frac{t}{2}-\frac{t^2}{8}\right)^2}}dt,
\end{equation}
\begin{equation}
    D_i(E) = \int \frac{t^3/16 - t^2/8  + Et/2 + f^2 - E}
            {\sqrt{2f^2t-\left(E+\frac{t}{2}-\frac{t^2}{8}\right)^2}}dt.
\end{equation}
The limits of the integration are the roots or the equation
$$
    2f^2t-\left(E+\frac{t}{2}-\frac{t^2}{8}\right)^2 = 0.
$$
For energies $E_2 < E < E_\mathrm{sep}$ 
corresponding to the classical region $2$, this equation has only two 
real roots.  For energies $E_\mathrm{sep} < E < E_1$ corresponding to regions 1 and 3, 
there are four real roots. 
$t_1 < t_2 < t_3 <t_4$. To obtain $T_1$, $K_1$ and $D_1$, the limits of integration 
should be $t_1$, $t_2$, and for $T_3$, $K_3$, $D_3$, they should be $t_3$, $t_4$.
Finally, for $E > E_1$, which corresponds only to region 3, there are two real roots 
once again.

\end{document}

%% file: arxiv_submission_190808.bbl
%